\documentclass[prl,twocolumn,showpacs,superscriptaddress,oneside]{revtex4}%
\usepackage{amssymb}
\usepackage{amsfonts}
\usepackage{graphicx}
\usepackage{epsfig}
\usepackage{amsmath}
\begin{document}
\title{Multiorbital effects on the transport and the superconducting fluctuations in LiFeAs}
\author{F. Rullier-Albenque}
\email{florence.albenque-rullier@cea.fr}
\affiliation{Service de Physique de l'Etat Condens\'e, Orme des Merisiers, CEA Saclay (CNRS URA 2464), 91191 Gif sur Yvette cedex, France}
\author{D. Colson}
\affiliation{Service de Physique de l'Etat Condens\'e, Orme des Merisiers, CEA Saclay (CNRS URA 2464), 91191 Gif sur Yvette cedex, France}
\author{A. Forget}
\affiliation{Service de Physique de l'Etat Condens\'e, Orme des Merisiers, CEA Saclay (CNRS URA 2464), 91191 Gif sur Yvette cedex, France}
\author{H. Alloul}
\affiliation{Laboratoire de Physique des Solides, UMR CNRS 8502, Universit\'e Paris Sud, 91405 Orsay, France}

\date{published in Physical Review Letters on November 2, 2012}

\begin{abstract}
Resistivity, Hall effect and transverse magnetoresistance (MR) have been
measured in low residual resistivity single crystals of LiFeAs. Comparison with ARPES and Quantum 
oscillation data implies that four  carrier bands unevenly contribute to the transport.
However the scattering rates of the carriers all display the $T^{2}$ behavior expected for 
a Fermi liquid. Near $T_{c}$ low field deviations of the MR with respect to a $H^{2}$
variation permit us to extract the superconducting fluctuation (SCF)
contribution to the conductivity. Though below $T_{c}$ the anisotropy of
superconductivity is rather small, the SCF display a quasi ideal two-dimensional
behavior which persists up to $1.4~T_{c}$. These results call for a refined theoretical understanding
of the multiband behavior of superconductivity in this pnictide.
\end{abstract}

\maketitle

\paragraph{Introduction.}
The superconductivity (SC) and normal state of the iron-based materials are
governed by their electronic structure involving the five iron 3$d$
orbitals, as established by band structure calculations and Angle Resolved
Photoemission (ARPES) studies \cite{Johnston}. A spin fluctuation
exchange mechanism for SC, with a $s^{\pm }$ symmetry, has been suggested
early on, based on the observation of a good nesting between hole and
electron Fermi surfaces \cite{Mazin}. This multiband character also highly
influences the  normal state transport and experimental studies, performed mainly in the
BaFe$_{2}$As$_{2}$(122) family, show that the electron mobility usually overcomes
that of holes \cite{FRA-BaFeCoAs, Kasahara-P}. But the conductivity $\sigma$ and Hall constant $R_{H}$
do not permit to disentangle the two carrier contributions to the
transport.

Unlike other iron-based SC, LiFeAs is a stoichiometric compound with
relatively high $T_{c}\simeq 18~$K without any chemical doping \cite{Tapp,
Pitcher}, so that it is a nearly compensated semi-metal, rather free from
defects. Cleaved single crystals display no surface states \cite{Lankau},
and clean ARPES data have evidenced \cite{Borisenko1, Umezawa} a dominant
large hole-like Fermi surface (FS) and a much smaller one centered at the $\Gamma$ point and two
electron-like FS at the corners of the Brillouin zone. 
The absence of matching between the sizes of the electron and hole FS's being a major difference with 
other pnictide families, it has been claimed \cite{Borisenko1} 
that the poor nesting prevents spin fluctuations driven SC,
while an orbital fluctuation mechanism \cite{Kontani} could be more appropriate.

In this letter, we take benefit of the reduced defect content in LiFeAs to take
accurate magnetoresistance (MR) data, which together with $\sigma$ and 
$R_{H}$ should permit to determine unambiguously the $T$ dependences of the
carrier contents and mobilities. While electron bands are found to dominate the
transport as in undoped BaFe$_2$As$_2$, the detailed quantitative comparison with ARPES 
\cite{Borisenko1,Umezawa} and de Haas-van-Alphen (dHvA) data \cite{Putzke} shows that
the holes involved in the largest hole band have the weakest mobility in LiFeAs. 
Furthermore, the MR data permits us a precise determination of the
superconducting fluctuations (SCF) contribution to the conductivity using
the method we recently established for the cuprates \cite{FRA-PRL2007, FRA-PRB2011}. These
SCFs have been so far poorly studied in multiband SC, even in MgB$_{2}$ 
\cite{Koshelev}, and  usually give information on the microscopic
properties of the SC state \cite{Larkin}. Here, we
find that the SCF paraconductivity can be very well fitted by the gaussian
Ginzburg Landau expectation for two dimensional single band SC systems. These results
should trigger theoretical studies of SCF taking into account both the
multiband aspect and the microscopic origin of SC, such as that initiated in
ref. \cite{Fanfarillo}.

\begin{figure}
\centering
\includegraphics[width=7cm]{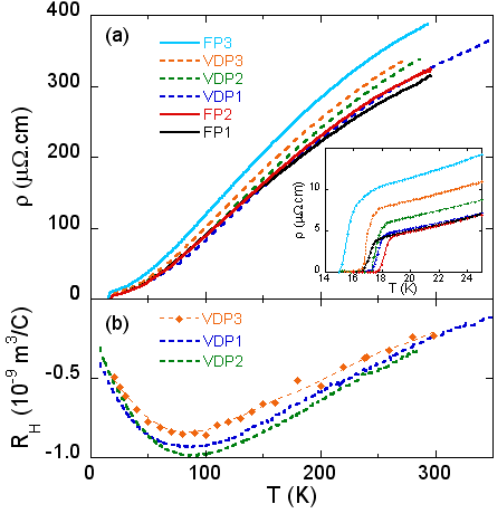}
\caption{(color on line) (a) In plane-resistivity versus temperature for the six samples of LiFeAs measured.
An expanded view of the superconducting transitions is reported in the inset. Except for one sample, 
one can see a correlation between the values of $T_c$ and those of the residual resistivity.
(b) Hall coefficient versus temperature for the three samples mounted in the Van der Pauw configuration.}
\label{Fig.rho(T)-R_H(T)}
\end{figure}

\paragraph{Resistivity and Hall effect} 
Six samples grown by a self flux technique as detailed in \cite{sup_mat} were studied, 
three with a four-probe configuration (labelled FP 1,2,3) and three with a Van der Pauw
configuration \cite{VdP} (labelled VDP1,2,3). The reproducibility of the
data for the in-plane $\rho (T)$ is displayed in Fig.\ref{Fig.rho(T)-R_H(T)}%
-a, the small differences (20\% at most at room $T$) being ascribed to
errors in the sample geometrical factors. The SC transition curves in the
inset of Fig.\ref{Fig.rho(T)-R_H(T)}-a evidence increasing $T_{c}$ values
(15.5 K to 18.2 K) for decreasing $\rho (T_{c})$, except for FP1. As shown below, 
the data can be fitted with $\rho(T)=\rho _{0}+AT^{2}$ 
below $\simeq 30$K. For our best samples FP1 and FP2, $\rho _{0}\sim 1.3\mu\Omega.cm$ corresponds to 
residual resistivity ratios $RRR=\rho(300$K$)/\rho_{0}\simeq 250$, five times larger than
previously reported \cite{Lee, Song, Heyer, Kasahara}. Comparing all those
data \cite{sup_mat}, we find that both $\rho_0$ and $A$
increase with decreasing $T_{c}$, \textit{i.e} Matthiessen's rule does not
apply and $\rho (T)$ cannot be analysed in \textit{a single-band model in
LiFeAs}. 

Whatever the temperature, we found that the Hall voltage is linear in field up to 14T. The similar 
negative Hall coefficients $R_{H}$, reported in Fig.\ref{Fig.rho(T)-R_H(T)}-b 
for three samples with slightly different $T_{c},$ show that electrons dominate the transport 
in this nearly compensated compound as in the nonmagnetic state of BaFe$_{2}$As$_{2}$ 
\cite{FRA-BaFeCoAs}. The broad minimum of $R_{H}(T)$, seen around 100K, coincides
with the observed change of curvature in the $\rho(T)$ curves. $R_{H}(T)$
tends towards zero with increasing $T$, which signals that the mobilities of
holes and electrons become similar. This behaviour bears some resemblance
with that found in overdoped Co-doped BaFe$_{2}$As$_{2}$ samples which also
exhibit a minimum around 100K, albeit less pronounced, and a similar
increase towards room $T$ \cite{FRA-BaFeCoAs}.

\paragraph{Transverse magnetoresistance.}

\begin{figure}
\centering
\includegraphics[width=7cm]{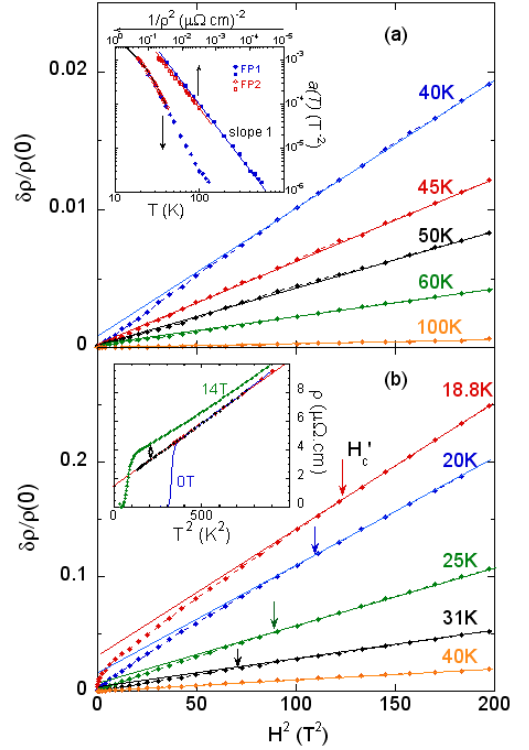}
\caption{(color on line) The transverse magnetoresistance measured in FP1 is plotted as a 
function of $H^2$ for $T>40$K in (a) and below 40K in (b). At high $T$, the MR is linear 
in $H^2$ whatever $H$ while it recovers this variation only above a given threshold field $H_c^{\prime} $
(indicated by arrows) for $T\lesssim 50$K. In the inset
of (a), the MR coefficient $a(T)$ of Eq.\ref{Eq.a(T)} is plotted versus $T$ and $1/\rho ^{2}$ 
for FP1 and FP2. The full lines of
slope 1 indicate that Kohler's rule is very well obeyed (see text). In the
inset of (b) the resistivity measured in 14 T field down to 10K is plotted
versus $T^{2}$ for FP1, together with the extrapolated zero field data.}
\label{Fig.MR}
\end{figure}

The small values of the residual resistivity $\rho_{0}$ permitted us to
perform accurate measurements of the transverse MR above $T_{c}$ up to $160$K. 
As shown in Fig.\ref{Fig.MR}a, for $T\gtrsim 45$K, the MR increases as $H^{2}$ 
in the whole field range investigated here ($H\leq 14$T). At lower $T$ (see Fig.\ref{Fig.MR}b), the low
$H$ increase of conductivity
detected when approaching $T_c$ is reminiscent of the contribution of superconducting fluctuations
as evidenced in YBCO \cite{FRA-PRB2011}.
These SCFs will be analysed later, but we shall focus first on the normal
state which is fully restored beyond a threshold field $H_{c}^{\prime }(T)$,
allowing us to define the MR coefficient $a(T)$ as 
\begin{equation}
\label{Eq.a(T)}
\delta\rho/\rho(T,0)=[\rho(T,H)-\rho(T,0)]/\rho(T,0)=a(T)H^2
\end{equation}
As shown in the inset of Fig.\ref{Fig.MR}a, $a(T)$ is found identical in FP1 and FP2 and to decrease by
about three orders of magnitude from $T_{c}$ to 160K. 

\paragraph{Compensated two-band model}

\begin{figure}
\centering
\includegraphics[width=8cm]{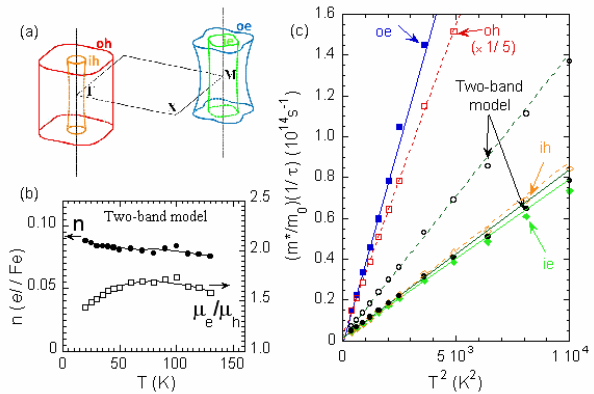}
\caption{(color on line) \textbf{a}: schematic view of the Fermi surfaces sheets of LiFeAs. \textbf{b}: number of carriers and mobility ratio extracted from the compensated two-band model. \textbf{c}: Scattering rates deduced in a two-band analysis (circles) and for the four band solution given in the text, plotted versus $T^2$. Full (empty) symbols and full (dashed) linear fits are for the electrons (holes). Notice that the data for the oh band has been divided by 5.}
\label{Fig.scatt-rates}
\end{figure}

The most natural approach to analyse these transport properties is within 
a two-band model. The transport coefficients are related to the respective conductivities 
$\sigma_h$, $\sigma_e$ and mobilities $\mu_h$, $\mu_e$ 
of the holes and electrons by:
\begin{eqnarray}
\rho^{-1} &=& \sigma=\sigma_e+\sigma_h \label{conductivity}\\
R_H &=& (\sigma_h\mu_h-\sigma_e\mu_e)/\sigma^2 \label{Hall}\\
\delta\rho/\rho(T,0)&=& \sigma_e\sigma_h(\mu_h+\mu_e)^2H^2/\sigma^2 \label{MR}
\end{eqnarray}
In undoped pnictides such as LiFeAs one expects equal number of electrons
and holes $n_{e}=n_{h}=n$ \cite{Borisenko1, Umezawa} and the two last
equations condense into $\sigma R_{H}=\mu_{h}-\mu_{e}$ and $a(T)=\mu_{e}\mu _{h}$,
so that the data permit us to deduce
the $T$ variations of  $\mu _{h}$, $\mu _{e}$ and $n$. In particular, with 
$R=1/ne$, these equations can be combined into  
\begin{equation}
a(T)=(R^{2}-R_{H}^{2})/4\rho^2(T,0).  \label{Kohler}
\end{equation}
We can see in fig.\ref{Fig.MR}-a that $a(T)$ scales as $\rho^{-2}$, that is
Kohler's rule \cite{Pippard} is well obeyed in LiFeAs, which indicates that 
$R^{2}-R_{H}^{2}$ has a weak $T$ dependence. We show in Fig.\ref{Fig.scatt-rates}c that 
both scattering rates display a $T^{2}$
variation up to $\sim 70$K \cite{sup_mat}, with $\mu_{e}/\mu _{h}\sim $1.5 (see Fig.\ref{Fig.scatt-rates}b) \cite{footnote}, 
and that the number of carriers is nearly $T$ independent below 120K. 
At higher $T$, thermal population of narrow bands might
induce an increase of $n(T)$ \cite{Heyer}, in analogy with the proposal we have done to explain
the transport properties of Ba(Fe$_{1-x}$Co$_x$)$_2$As$_2$ \cite{FRA-BaFeCoAs}. 

\paragraph{Comparison with ARPES and dHvA: beyond the two-band model}
However, the deduced carrier content $n \sim 0.09$ \textit{el}/Fe being twice smaller than that 
given by ARPES or by dHvA \cite{Umezawa, Putzke}, this implies that the two-band model only allows 
us to determine separate average parameters for the holes and electrons. Indeed, we do know from ARPES 
data \cite{Borisenko1, Umezawa} that hole carriers are located in a small
inner ($ih$) band included in a large outer ($oh$) band. Similarly, it has
been suggested \cite{Ferber} that, due to strong spin-orbit coupling, the
electron bands should not be considered as two crossed degenerate elliptic
pockets. As found in dHvA experiments in both LiFeP and
LiFeAs \cite{Putzke}, they split into an inner ($ie$) band included in an
outer one ($oe$). The carrier mobilities are
expected to differ substantially for these four bands, and the MR coefficient has now a more
complicated expression than given in eq.\ref{MR} \cite{Kuo}: 
\begin{equation}
H^{-2}\Delta\rho/\rho(T,0) = \sigma_e\sigma_h(\mu_e+\mu_h)^2/\sigma^2+\sigma_hA_h/\sigma +\sigma_eA_e/\sigma \label{MR-2}
\end{equation}
with $A_h = \sigma_{ih}\sigma_{oh}(\mu_{ih}-\mu_{oh})^2/\sigma_h^2$, where $\mu_h$ is now an effective
hole band mobility given by $\sigma_h\mu_h = \sigma_{ih}\mu_{ih}+\sigma_{oh}\mu_{oh}$ and an effective 
number of holes is defined by $\sigma_h=n_h^{eff}e\mu_h$. Similar expressions hold for the electron bands.

To  approach a solution we may use for the $T$ independent carrier contents those obtained from ARPES 
and dHvA experiments $n_{oh}\sim 0.16$ \textit{h}/Fe, $n_{ih}\sim 0.03$ \textit{h}/Fe,
$n_{oe}\sim0.11$ \textit{el}/Fe and $n_{ie}\sim 0.08$ \textit{el}/Fe 
for the outer and inner hole (electron) bands respectively. As we are still lacking sufficient experimental 
information, a unique solution cannot be acquired. However dHvA results imply that the
mobilities for the two electron bands are comparable. Matching the data with all these assumptions yields
a strong differentiation of the two hole bands with a surprisingly much lower mobility for the outer 
band compared to the inner one. Also, imposing
$\mu_{oe}\simeq\mu_{ie}$ puts some constraints on the value of $n_{h}^{eff}$ that cannot be larger
than $\sim 0.06$ \textit{h}/Fe \cite{sup_mat}. 
A solution with $n_{h}^{eff}=0.05$ \textit{h}/Fe whatever $T$ is illustrated in 
Fig.\ref{Fig.scatt-rates}c and gives a ratio between the electron (hole) mobilities respectively 
of $\mu _{ie}/\mu_{oe}=3-6$ and $\mu _{ih}/\mu _{oh}\sim 17$. It corresponds to 
$n_{e}^{eff}\sim 0.13$ \textit{el}/Fe
and it results in similar values of the effective electron and
hole mobilities \cite{sup_mat}, which would justify 
why Kohler's rule is obeyed in this compound. Whatever the value taken for $n_h^{eff}$ in 
the range considered, we always find that the
scattering rates for the different carriers increase as $T^{2}$ up to $\sim 70$K, as seen in 
Fig.\ref{Fig.scatt-rates}c. This
confirms a Fermi liquid behavior for both holes and electrons in agreement with density 
functional theory and dynamical mean Field (DFT+DMFT) calculations \cite{Ferber}.

The $T=0$ extrapolated values correspond to mobilities 
$\sim1000$ $cm^2$/Vs for the smallest hole and electron bands, that is, using a Fermi velocity 
$v_{F}\sim 1~10^{5}$ m/s \cite{Kordyuk} and an effective mass
$m^*\simeq4m_0$ \cite{Putzke}, to mean free paths $\approx$ 2000 \AA , in agreement with
the high purity of our FP1 sample. Much faster relaxation is observed for the outer hole band which 
is exclusively constructed from the $d_{xy}$ orbital, for which a stronger
incidence of correlations on $m^*$ and the scattering rates is expected from DFT+DMFT calculations 
\cite{Ferber,Yin}. 

\paragraph{Superconducting fluctuation contribution to conductivity.}
We detail in \cite{sup_mat} why the low field deviations from a $H^2$ behavior of the MR cannot be
associated with a saturation of the normal state MR. This is confirmed experimentally, as such deviations
were not detected in LiFeP, which has a lower $T_c=7$K than LiFeAs, with a similar band structure and
residual resistivity as our samples \cite{Kasahara}. Following our extensive study done in cuprates
\cite{FRA-PRB2011}, the SCF contribution to the conductivity is given by
$\Delta \sigma _{SF}(T)=\rho ^{-1}(T)-\rho _{n}^{-1}(T)$, 
where $\rho_n(T)$ is the $H=0$ extrapolation of the $H^{2}$
variation of the normal state MR. Let us notice here that $\Delta \sigma _{SF}(T)$, which does not
exceed $3$\% of the normal state values (see fig.\ref{Fig.MR}b), 
would be extremely difficult to extract directly from the $\rho(T)$ curves.  

\begin{figure}
\centering
\includegraphics[width=7cm]{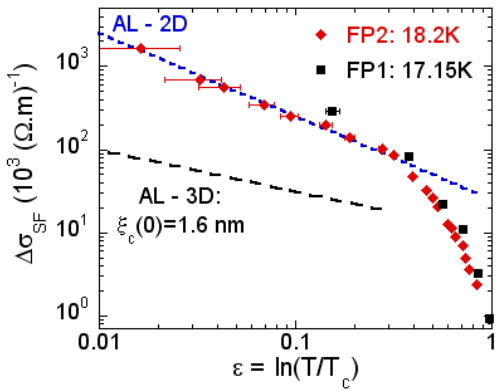}
\caption{(color on line)Superconducting fluctuation conductivity $\Delta 
\protect\sigma _{SF}$ for samples FP1 and FP2 as a function of $\protect%
\epsilon =\ln (T/T_{c0}),$ where $T_{c0}$ is taken at the midpoint of the
transition. The error bars for $\epsilon$ were determined using $T_c$ values at 25\% and 75\% of the
resistive transition. Estimates of the AL contribution in dimensions 2 and 3 are displayed as dashed lines for comparison.}
\label{Fig.fluct}
\end{figure}

For our samples with the highest $T_{c0}$, the $\Delta \sigma _{SF}$ data reported in a log-log scale 
in Fig.\ref{Fig.fluct} as a function of the reduced temperature $\epsilon =\ln(T_c/T_{c0})$ resemble
those found in cuprates \cite{FRA-PRB2011}. After an initial power law behavior, $\Delta \sigma _{SF}(T)$
displays a cut-off, and only becomes negligible for $\epsilon\sim1$.
This larger SCF regime $\sim 2.5~T_{c}$ than that we did find in cuprates points towards a 2D character. 
This is better seen from the good fit with the 2D Aslamazov-Larkin formula
\begin{equation}
\Delta \sigma ^{AL}(T)=(e^{2}/16\hslash s)~\epsilon^{-1},  \label{AL}
\end{equation}
for $0.017\lesssim \epsilon \lesssim 0.3$,
corresponding to $T=1.02$ to 1.4$T_{c}$, without any fitting parameter
except the interlayer distance $s$ taken here as  the lattice
parameter value $c=6.36$\AA. The value of $T_c$ is taken here at the midpoint of the transition but our conclusion remains valid for slightly different $T_c$ determinations as shown by the error bars for $\epsilon$ in Fig.\ref{Fig.fluct}. 

These 2D SCFs appear at first sight difficult to reconcile with the 3D character
of the normal and SC states in LiFeAs suggested by the weak anisotropies of resistivity and upper-critical
field $H_{c2}$ \cite{Song, Kurita, Cho} (see also \cite{sup_mat}).
Using for instance a single-band scheme to deduce the coherence length along the $c$-axis from $H_{c2}$,
we find $\xi_c\sim$16 \AA, that is more than twice the
interlayer spacing. So one would expect for 3D fluctuations the much smaller contribution to the
conductivity shown in Fig.\ref{Fig.fluct}.
All this suggests that the present result is specific to
multiband effects and that, above $T_c$, the fluctuating pairs are driven by a single 2D band.
In view of our discussion above on the normal state, only the outer hole pocket that originates 
from in-plane $d_{xy}$ orbitals is purely 2D\cite{Ferber, Yin, Borisenko2}.

It is worth comparing our results in LiFeAs with the case of MgB$_2$, a presumably much simpler multiband
SC which also displays a rather low $H_{c2}$ anisotropy factor $\sim2$ at $T_c$. It has been proposed
\cite{Koshelev} that the SCFs are governed by a unique critical mode, which is dominated for 
$T>>T_c$ by the quasi 2D $\sigma$ bands with the larger SC gap. However, near $T_c$, 
the critical mode should recover a 3D character due to both band contributions and 
the paraconductivity should diverge slightly slower 
than $\sim1/\sqrt{\epsilon}$ . Preliminary data \cite{Sidorenko} have 
been analysed as suggesting 2D SCFs but, to our knowwledge, no more reliable experimental work has been performed since.

Concerning the properties of the SC gaps in LiFeAs, they are found weakly anisotropic by ARPES, the largest 
$\Delta =5-6$meV for the small 3D hole pocket, while $\Delta =3-4$meV for the other bands 
\cite{Umezawa, Borisenko2}, including the large 2D hole band. In that context one should rather expect 
a pronounced 3D character for the SCFs in LiFeAs \cite{Lara}.

\paragraph{Discussion.}

In the specific case of pnictides, it has been underlined recently \cite{Fanfarillo} that the SCFs
are indeed expected to behave differently as, contrary to the case of MgB$_2$ \cite{Koshelev}, the pairing
should be dominated by interband spin fluctuation interactions \cite{Qureshi}. 
These authors have shown that, in such a case, the critical mode controlling the SCFs should have
a simpler relation to the various bands, so that the AL formula known for single band SC should 
remain valid either in 2D or 3D. Further experimental investigations of the
SCFs along the same lines in other pnictide families would be necessary to get a more complete 
understanding of SCFs in these multiband compounds. For instance, one needs to ascertain whether the SCFs 
are dominated  by a 3D contribution in the hole doped BaKFe$_2$As$_2$, as has been suggested from 
an investigation of the diamagnetism above $T_c$ \cite{Mosqueira}.

To conclude, we emphasized here the relevance of MR data to unveil new physical phenomena. Our data give
indications which should help to investigate the incidence of spin fluctuations on the carrier scattering 
in the normal state and on the SC pairing in this specific LiFeAs system. On the other hand,
it would be interesting to study whether other proposals such as orbital fluctuations or p-wave SC
\cite{Kontani,Hanke} might provide a natural explanation for these 2D SCFs. 
Unveiling the origin of these 2D SCFs should therefore give useful hints to clarify the 
mechanism of SC in LiFeAs .

\begin{acknowledgments}
This work has been performed within the "Triangle de la Physique" and was supported by ANR grant "PNICTIDES".
We thank L. Benfatto, R. Valenti and A. Coldea for helpful scientific exchanges about the interesting specifics of this LiFeAs compound.
\end{acknowledgments}

\pagebreak 

\section{Supplementary material}
\paragraph{Samples and transport measurements}
Single crystals of LiFeAs were grown using a self-flux method. Lumps of Li and powders of FeAs and As in 
the molar ratio 3:2:3 were placed in an alumina crucible in an argon glove box. The crucible was inserted
into a ceramic container covered by a cap and sealed in an evacuated quartz tube. The samples were heated 
to 1090$^\circ$C with a rate of 175$^\circ$C/h, held at this temperature for 4h then cooled down to
800$^\circ$C at 6$^\circ$C/h and then more rapidly to room temperature. Clean crystals of typical 
dimensions $5$x$5$x$0.05$mm$^3$ were mechanically extracted from the flux. A Fe:As ratio of 1:1 has been
confirmed by electron microprobe. 

Six different
samples from the same batch with different geometries were studied. They were cleaved from larger crystals 
to thicknesses ranging from 18 to $60\mu$m. Electrical 
contacts were made using silver epoxy. Owing to the very fast degradation of the surfaces under air 
exposure, the samples were covered with Apiezon N-grease immediately after putting the contacts and 
very rapidly inserted in the cryostat. 

Hall effect measurements were performed in the Van der Pauw
configuration \cite{VdP} for three samples labelled VDP1,2,3 while standard 4-probe technique was 
used for magnetoresistance measurements in three other samples (labelled FP1,2,3). 
The Hall effect was measured  by exchanging 
the voltage and current probes in the diagonal directions of the VdP samples either in a fixed 14T magnetic
field (VDP1 and 2) or by sweeping the magnetic field up to 14T (VDP3). In all cases, it was checked that the Hall voltage increases linearly with magnetic field, allowing us to determine unambiguously the Hall coefficient. The transverse magnetoresistance (MR) has 
been measured on the samples FP1 and 2 at fixed temperature by sweeping the magnetic field from -14 to 
14T and taking the symmetric part of the signal in order to eliminate any spurious component due to
misalignment of the contacts. As the MR becomes rather small above $T>100$K (usually less than 0.05\% 
at 14T) great care has been taken to ensure that the temperature remains constant during the magnetic 
field ramp. We have used a cernox sensor which has been calibrated in magnetic field and the temperature 
was then regulated by compensating the incidence of the field on the sensor value.

\paragraph{Characteristics of the different samples}
The transport and superconducting parameters found for the different samples studied here are gathered in Table \ref{tab.1} and compared to some data published in the literature.
\begin{table}[h]
\caption{The transport parameters for the different samples studied are compared with those reported 
in ref.\cite{Heyer, Kasahara}.The values of $T_c$ are taken at the mid-point of the resistive transition. 
The residual resistivities $\rho_0$ have been determined by fitting the zero field $\rho(T)$ curves 
with a $T^{2}$ dependence $\rho(T)=\rho_0+AT^2$. The residual resistivity ratio RRR is given by $\rho(300K)/\rho_0$.}
\label{tab.1}
\begin{center}
\begin{tabular}{ccccccc}
\textbf{Sample} & $T_{c}$ & $\rho_0$ & RRR & $A$ \\
  & (K) & ($\mu\Omega.cm$) &   & ($n\Omega.cm/K^2$)\\
VDP1 & 17.75 & 4.15 & 80 & 9.7\\
VDP2 & 17.6 & 2.93 & 119 & 9.3\\
VDP3 & 17 & 4.55 & 79 & 9.5\\
FP1 & 18.2 & 1.47 & 225 & 8.8\\
FP2 & 17.15 & 1.21 & 263 & 9.3\\
FP3 & 15.6 & 6.34 & 62 & 12.7\\
Ref.\cite{Heyer} & 17.75 & 15.2 & 38 & 22\\
Ref.\cite{Kasahara} & 17.3 & $13$ & 53 & 20\\
\end{tabular}
\end{center}
\end{table}
Diamagnetic shielding was also measured for sample VDP1 and gave an onset $T_c$ value of 17.4K with 
a transition width of 2K, in close agreement with the transport measurements.
One can see that the coefficient $A$ is strongly dependent on the purity of the sample. This means that,
as could be expected for a multiband compound, 
Matthiessen's rule is not obeyed. Consequently, the
comparison between the values of $A$ and the electronic specific heat coefficient through the 
Kadowaki-Woods ratio, which is used to estimate the strength of electronic correlations in 
single-band metals, cannot be done so straightforwardly here.

\paragraph{Resolution of the problem beyond the two-band model}
As explained in the text, we have to solve a problem with four unknown mobilities and only three 
experimental parameters. From dHvA results we know that the mobilities for the two electron bands are 
similar and must be larger than those of the hole bands. 
If we consider the limit where they are equal, that is $\mu_{ie}=\mu_{oe}$ with
$n_e^{eff}=n_{ie}+n_{oe}$, we find that the mobility of the inner hole band is much larger than that
of the outer one, and also larger than the electron mobility. In that case, the effective 
number of holes defined as $\sigma_h=n_h^{eff}\mu_h$ is equal to $n_{ie}$ taken equal here to 0.03\textit{h}/Fe. 

A more physical solution can thus be obtained by increasing $n_h^{eff}$, which consequently
lets the hole outer band participate to the transport and slightly differentiates the two electron
bands, with $\mu_{oe}\lesssim\mu_{ie}$. However if $n_h^{eff}$ is taken too large, typically 
$\gtrsim 0.06$\textit{h}/Fe, the mobilities for the two electron bands become very different, 
in contradiction with the dHvA results. 
Consequently $0.03<n_h^{eff}<0.06$ is required to match the transport data
together with the ARPES and dHvA results. The solution given in the text is obtained for 
$n_h^{eff} =0.05$\textit{h}/Fe. At this stage we are not able to better refine 
the resolution of the problem and $n_h^{eff}$ is taken $T$ independent, which means 
that the ratio between the mobilities of the two hole bands is assumed to remain constant with temperature. 
Though this is not justified, this hypothesis should not distort the main conclusions drawn 
from this analysis.

\begin{figure}
\centering
\includegraphics[width=8cm]{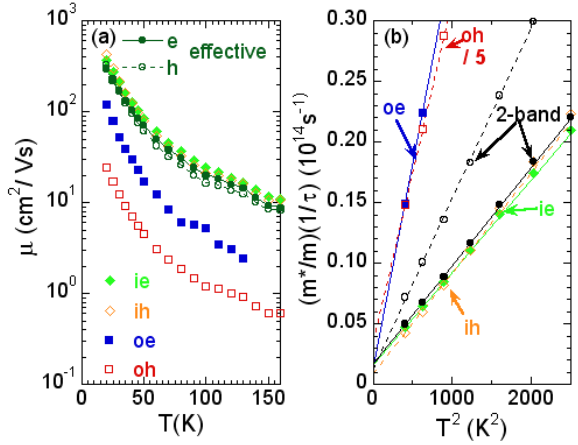}
\caption{(color on line) (a): electron and hole mobilities for the four different bands by using
a realistic model differentiating the four bands as explained in the text. The effective mobilities for the electron and hole bands are also plotted and found nearly identical to the larger mobilities of the inner bands.
(b) Scattering rates weighted by their effective masses plotted versus $T^2$ for the electrons and holes, using either a compensated two band model or the four band analysis considered in a. Lines are linear fitting curves.}
\label{Fig_mobility}
\end{figure}

Technically, the resolution of the problem then consists in extracting the different mobilities of 
the electrons and the effective mobility of the holes from Eqs. 2, 3 and 6 of the text. 
The values of the number of holes in the two different bands ($n_{ih}=0.03$, $n_{oh}=0.16$) and 
$n_h^{eff}=0.05$ imply $\mu_{oh}\ll\mu_{ih}$ and the term $A_h$ in Eq.6 can be neglected. 
The second term $\sigma_eA_e/\sigma$ represents at most $\sim25$\% of the total MR coefficient. 
Only one solution is found at each temperature, giving the different mobilities plotted in fig.\ref{Fig_mobility}a. 
One can see that the effective mobilities found for the holes and electrons are nearly equal. This explains
why Kohler's rule is well obeyed in this compound. Indeed in a two-band model, this rule directly 
follows from the fact that the two different types of carriers have the same mobility.

The low temperature behavior of the corresponding scattering rates weighted by the effective mass 
is displayed in Fig.\ref{Fig_mobility}-b where goods fits with a $T^2$ law is found
up to $\sim 70$K. Although the actual $T=0$ extrapolated values has to be considered with caution as they
depend on the chosen value of $n_h^{eff}$, we have checked that we always find comparable values of the 
order of 2 $10^{12}$s$^{-1}$ for all the carriers except for the outer hole band. 
This would yield mean free paths for the electrons about 5 times larger than those found in dHvA 
experiments \cite{Amalia} and thus let us foresee that the inner hole band should be also observed by dHvA 
in our cleanest samples.

\paragraph{Possible deviations of the MR from a $H^2$ behavior} 
The expressions (Eqs. 3 and 4) given in the text for the Hall coefficient and the transverse MR
are the first order approximation in $H^2$. At a higher order, we get:
\begin{eqnarray}
\rho_{xy}(H) &=& \frac{(\mu_h\sigma_h-\mu_e\sigma_e)H}{(\sigma_h+\sigma_e)^2+(\sigma_e\mu_h-\sigma_h\mu_e)^2H^2} \label{Hall-sat}\\
\frac{\delta\rho}{\rho(0)} &=& \frac{\sigma_e\sigma_h(\mu_e+\mu_h)^2H^2}{(\sigma_h+\sigma_e)^2+(\sigma_e\mu_h-\sigma_h\mu_e)^2H^2} \label{MR-sat}
\end{eqnarray}
This can introduce some saturation of the Hall coefficient and MR when the second term in the denominator
$(\sigma_e\mu_h-\sigma_h\mu_e)^2H^2$ becomes comparable or larger than $\sigma^2$.
For a compensated material with $n_e=n_h$ this term is always zero and no saturation of the Hall 
resistivity nor magnetoresistance is expected with increasing magnetic field. 
Deviations from a $H^{2}$ behavior can in principle occur if $n_e \neq n_h$.
However, this is unlikely here as we found that the Hall effect that
should display the same effect remains linear in $H$ within a 5 10$^{-3}$ accuracy in the same
temperature range. Moreover, using the values of the effective mobilities found above, we
expect at most a change of slope of $\sim 4$\% for the MR at 20K between 0 and 14T,
while the observed deviation corresponds to nearly a factor 2. Consequently the downward shifts 
observed at low magnetic field can only arise from the contribution of superconducting fluctuations 
and its suppression by the magnetic field.

\begin{figure}
\centering
\includegraphics[width=8cm]{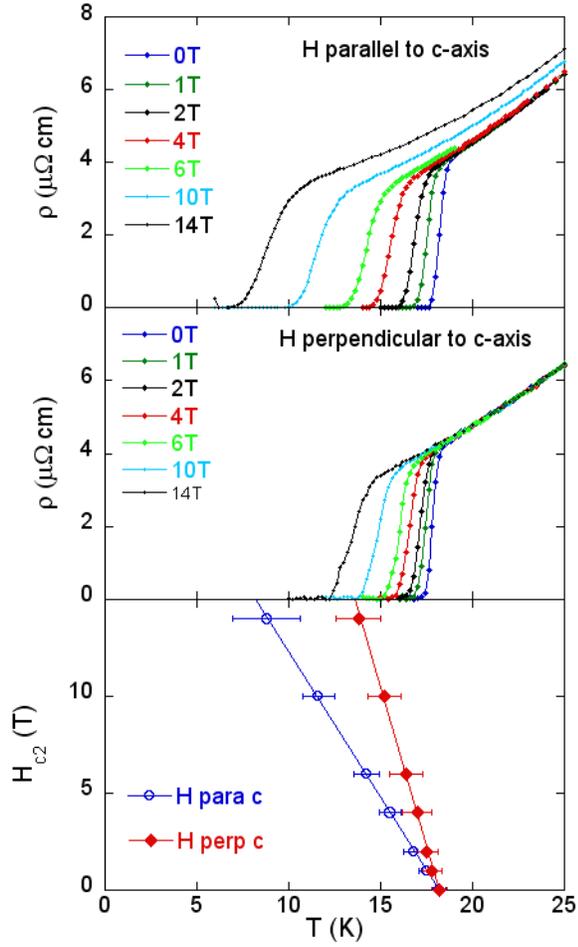}
\caption{(color on line)Resistive transitions of the sample FP2 at different magnetic fields applied 
parallel (a) and perpendicular (b) to the c-axis. (c) Upper-critical fields $H_{c2}(T)$ for the 
magnetic field applied parallel (empty circles) and perpendicular (full diamonds) to the c-axis. 
The error bars represent the transition widths measured at 10-90\% of the resistive transitions.}
\label{fig_rho(T)-fields}
\end{figure}

\paragraph{Determination of the upper-critical fields}
Fig.\ref{fig_rho(T)-fields} shows the resistive superconducting transitions of the sample FP2 
for different magnetic fields up to 14T parallel (a) or perpendicular (b) to the c-axis. Only a small 
broadening of the SC transitions is observed with increasing $H$, allowing us to define the critical
temperature $T_c(H)$ at the mid-point of the resistive transition for each magnetic field quite precisely. 
For both orientations of the magnetic field, the increase of $H_{c2}$ with decreasing $T$ shows a linear 
dependence as shown by the full lines in Fig.\ref{fig_rho(T)-fields}c with a ratio between the slopes 
of $\sim2.2$ in excellent agreement to what was found in earlier reports \cite{Heyer, Kasahara}.


\begin{thebibliography}{99}
\bibitem{footnote}The values of the electron and hole mobilities, and their ratio, determined here are very different from those reported in ref.\cite{Kasahara} in which $\mu_e/\mu_h \sim10$. This comes from numerical errors in their determination of mobilities from their experimental data. The correct estimate $\mu_e/\mu_h \sim1.4$ is very similar to ours.
\end{thebibliography}
\end{document}